\documentclass[onecolumn,prd,superscriptaddress,showpacs,floatfix,%
preprintnumbers,nofootinbib,showkeys]{revtex4}  

\usepackage{axodraw}
\usepackage{amsmath,amsfonts}
\usepackage{epsfig,epsf}

\def\be {\begin{equation}}
\def\ee {\end{equation}}
\def\beq {\begin{equation}}
\def\eeq {\end{equation}}
\def\bea {\begin{eqnarray}}
\def\eea {\end{eqnarray}}
\def\barr{\begin{array}}
\def\earr{\end{array}}

\def\opcit(#1){ {\em op. cit.}, #1}

\def\issue(#1,#2,#3){#1, #2 (#3)} 

\def\APP(#1,#2,#3){Acta Phys.\ Polon.\ \issue(#1,#2,#3)}
\def\ARNPS(#1,#2,#3){Ann.\ Rev.\ Nucl.\ Part.\ Sci.\ \issue(#1,#2,#3)}
\def\CPC(#1,#2,#3){Comp.\ Phys.\ Comm.\ \issue(#1,#2,#3)}
\def\CIP(#1,#2,#3){Comput.\ Phys.\ \issue(#1,#2,#3)}
\def\EPJC(#1,#2,#3){Eur.\ Phys.\ J.\ C\ \issue(#1,#2,#3)}
\def\EPJD(#1,#2,#3){Eur.\ Phys.\ J. Direct\ C\ \issue(#1,#2,#3)}
\def\IEEETNS(#1,#2,#3){IEEE Trans.\ Nucl.\ Sci.\ \issue(#1,#2,#3)}
\def\IJMP(#1,#2,#3){Int.\ J.\ Mod.\ Phys. \issue(#1,#2,#3)}
\def\JHEP(#1,#2,#3){J.\ High Energy Physics \issue(#1,#2,#3)}
\def\JPG(#1,#2,#3){J.\ Phys.\ G \issue(#1,#2,#3)}
\def\MPL(#1,#2,#3){Mod.\ Phys.\ Lett.\ \issue(#1,#2,#3)}
\def\NP(#1,#2,#3){Nucl.\ Phys.\ \issue(#1,#2,#3)}
\def\NIM(#1,#2,#3){Nucl.\ Instrum.\ Meth.\ \issue(#1,#2,#3)}
\def\PL(#1,#2,#3){Phys.\ Lett.\ \issue(#1,#2,#3)}
\def\PRD(#1,#2,#3){Phys.\ Rev.\ D \issue(#1,#2,#3)}
\def\PRL(#1,#2,#3){Phys.\ Rev.\ Lett.\ \issue(#1,#2,#3)}
\def\SJNP(#1,#2,#3){Sov.\ J. Nucl.\ Phys.\ \issue(#1,#2,#3)}
\def\ZPC(#1,#2,#3){Zeit.\ Phys.\ C \issue(#1,#2,#3)}

\usepackage[dvips]{color}
\usepackage[normalem]{ulem}
\definecolor{darkgreen}{cmyk}{1,0,1,0.4}

\definecolor{darkred}{cmyk}{0,1,1,0.4}

\usepackage{slashed}

\newcommand{\kay}{\slashed k}

\newcommand{\pone}{\slashed p_1}

\newcommand{\mf}{m_f}
\newcommand{\vf}{v_f}
\newcommand{\af}{a_f}
\newcommand{\vb}{v_b}
\newcommand{\ab}{a_b}
\newcommand{\vzf}{v_Z^f}
\newcommand{\azf}{a_Z^f}
\newcommand{\vzb}{v_Z^b}
\newcommand{\azb}{a_Z^b}
\newcommand{\aplus}{{\cal A}_+}
\newcommand{\aminus}{{\cal A}_-}

\begin{document}

\title{
Z-pole observables in an effective theory}

\author{Debajyoti Choudhury}
\email{debajyoti.choudhury@gmail.com}
\affiliation{Department of Physics and Astrophysics, 
University of Delhi, Delhi 110007, India.}

\author{Anirban Kundu}
\email{anirban.kundu.cu@gmail.com}
\affiliation{Department of Physics, University of Calcutta,\\
92, Acharya Prafulla Chandra Road, Kolkata 700009, India. }

\author{Pratishruti Saha}
\email{pratishruti.saha@umontreal.ca}
\affiliation{Physique des Particules, Universit\'{e} de Montr\'{e}al, 
Case Postale 6128, succursale centre-ville, Montr\'{e}al, QC, 
Canada H3C 3J7.}

\date{\today}

\begin{abstract} 

There are two Z-peak observables related to the pair production of 
bottom quarks that show a deviation of about $2.5 \sigma$ each
from Standard Model expectations. While the discrepancy in the
forward-backward asymmetry is a long-standing one, the tension for 
the second observable, namely the ratio of the partial width for a Z 
decaying to a pair of bottom quarks to the total hadronic decay
width of the Z, has recently gone up due to a full two-loop
evaluation of the Standard Model contributions. 
We show how both these discrepancies may be explained in the framework 
of new physics that couples only to the third generation of quarks.
In the paradigm of effective operators, the Wilson coefficients of
some of the possible operators are already very tightly constrained
by flavour physics data. However, there still remain certain 
operators, particularly those involving right-chiral quark fields,
which can successfully explain the anomalies.  We also show how
the footprints of such operators may be observed at the upgraded
LHC.
\end{abstract}


\maketitle

\section{Introduction}
\label{sec:introduction}

The majority of the electroweak precision observables are in good 
agreement with the Standard Model (SM)~\cite{gfitter}. However, there
are two which show a marked tension, albeit not at the level where 
they can be claimed as incontrovertible evidence for New Physics (NP)
beyond the SM. One of these is the long-standing anomaly of
forward-backward asymmetry in the pair-production of $b$ quarks,
$A^b_{FB}$, as measured at the $Z$-peak. The second is the ratio 
$R_b$, defined as 
$R_b = \Gamma (Z \to b\bar{b})/\Gamma (Z \to {\rm hadrons})$. 
Of much interest during the LEP-I and SLC 
era~\cite{LEPEW:1994aa,Abe:1995iq,Abe:1999hb,ALEPH:2005ab}, 
the second tension has resurfaced due to a recent evaluation of $R_b$ 
in the SM, taking into account all two-loop effects~\cite{rb-theory}. 

The Gfitter group~\cite{gfitter} has updated the SM fit after the 
discovery of the Higgs boson at 
$m_h = 125.7\pm 0.4$ GeV~\cite{cms-higgs,atlas-higgs}. 
With the experimental
inputs from Ref.~\cite{lepewwg}, the fit~\cite{gfitter} shows
\be
R_b~({\rm exp}) = 0.21629\pm 0.00066\,,\ \ 
R_b~({\rm SM}) = 0.21474\pm 0.00003\,,
\label{eq:R_b-data}
\ee
with a pull of $-2.35$, where for any observable $O$ with a standard 
deviation $\sigma_{\rm exp}$, the pull is defined as\footnote{This
definition is consistent with Gfitter but opposite in sign to that used
by the Particle Data Group (PDG)~\cite{pdg}.}
\be
{\rm Pull} = \frac{O_{\rm SM} - O_{\rm exp}}{\sigma_{\rm exp}}\,.
\label{eq:def-pull}
\ee
Note that the pull has increased to $-2.35$ from $-0.8$ (as calculated earlier
using $R_b({\rm SM}) = 0.21576\pm 0.00008$) thanks to the recent
computation of the full 2-loop effects in the SM~\cite{rb-theory}. 

The pull for $A^b_{FB}$ is $2.5$, computed from\cite{lepewwg,gfitter} 
\be
A^b_{FB}~({\rm exp}) = 0.0992\pm 0.0016\,,\ \ 
A^b_{FB}~({\rm SM}) = 0.1032^{+0.0004}_{-0.0006}\,.
\label{eq:AFB_b-data}
\ee
Present ever since the LEP-I days, this discrepancy constitutes,
perhaps, the most longstanding indicator of NP. 
Indeed, over the years, numerous attempts have been made to solve
this problem in the context of specific NP scenarios. Prominent
amongst these are those invoking 
extra Higgs scalars~\cite{Haber:1999zh}, low energy
supersymmetry~\cite{Altarelli:2001wx} or just mixing with exotic
quarks~\cite{Choudhury:2001hs}. 
At the same time, the data has a significant constraining power
and may be used to rule out certain classes of models particularly in
the light of the discovery of the 126 GeV scalar\footnote{For
example, no supersymmetric model, where the lighter chargino is
dominantly a wino, is consistent with both $R_b$ and $A^b_{FB}$
measurements, if we assume the 126 GeV scalar to be the lightest 
CP-even neutral Higgs boson~\cite{ggaktsr}.}.

There is another mild tension in the forward-backward asymmetry of the
$\tau$ measured at the $Z$-peak. While this has not been updated
in Ref.~\cite{gfitter} using the $m_h$ data, the value of the asymmetry
hardly depends on whether $m_h$ is given as an input or is treated as 
a free parameter to be determined from the fit. We therefore quote 
the PDG result~\cite{pdg}:
\be
A^\tau_{FB} ~({\rm exp}) = 0.0188\pm 0.0017\,,\ \ 
A^\tau_{FB} ~({\rm SM}) = 0.01633\pm 0.00021\,,
\label{eq:AFB_tau-data}
\ee
with a pull of $-1.5$. However, the branching ratio for
$Z\to\tau^+\tau^-$ is consistent with that of the other leptons, viz.
\be
Br(Z\to\tau^+\tau^-) = (3.370\pm 0.008)\%\,,\ \ 
Br(Z\to e^+e^-) = (3.363\pm 0.004) \%\,.
\label{eq:Zbf-data}
\ee
Taking into account the electroweak corrections, 
$R_\tau \equiv \Gamma(Z\to {\rm hadrons})/\Gamma(Z\to\tau^+\tau^-)$
is slightly above the SM predictions, but consistent nevertheless,
with a pull of only $0.6$:
\be
R_\tau~({\rm exp})= 20.764\pm 0.045\,,\ \ \ 
R_\tau~({\rm SM}) = 20.789\pm 0.011\,.
\label{eq:R_tau-data}
\ee

The partial width  $\Gamma(Z \to b \bar b)$ is best 
analysed by parametrizing the $Zb\bar{b}$ vertex as
\be
\frac{g}{\cos\theta_W} 
\bar{b} \gamma^\mu 
\left[(g_L^b+\delta g_L^b)P_L + (g_R^b + \delta g_R^b)P_R\right]
b Z_\mu 
\label{eq:Z_coup}
\ee
where 
\be
g_L^b = T_3^b - \kappa_b Q_b \sin^2\theta_W\,,\ \
g_R^b = - \kappa_b Q_b \sin^2\theta_W\,,
\label{eq:def-gLgR}
\ee
with $\kappa_b = 1.0067$. The deviation of $\kappa_b$ from unity incorporates 
the electroweak corrections, whereas $\delta g^b_{L,R}$ comprise all possible
corrections arising from NP sources. On analyzing all the electroweak
data\footnote{It should be noted that had we concentrated only on
  $\Gamma(Z \to b \bar b)$ and $A^b_{FB}$, to the exclusion of all
  else, the fit would have been substantially different, with $\delta
  g^b_L \sim 0.003$.  This, however, would be illogical for such a
  simple-minded shift would cause the predictions for several other
  precision variables (such as $\Gamma_Z$, $\Gamma_{\rm had}$ etc.) to
  deviate from the measurements.}, the best fits are obtained~\cite{batell} 
for
\beq
\barr{r crcl c rcr}
(i) & \quad & 
   \delta g_L^b & = & 0.001\pm 0.001 & \qquad & \delta g_R^b & = &  0.016\pm 0.005
\\[1.5ex]
(ii) & \quad & 
   \delta g_L^b & = & 0.001\pm 0.001 & \qquad & \delta g_R^b & = & -0.170\pm 0.005 
\earr
     \label{gl_gr_fit}
\eeq
where both $\delta g_L^b$ and $\delta g_R^b$ have been treated as free
parameters.  Indeed, the $\chi^2/d.o.f.$ for the two fits 
are too close to be called apart~\cite{batell,Choudhury:2001hs}.
It is  easy to see the origin of these two
solutions.  Apart from some numerical constants,
\beq
\Gamma(Z\to b\bar{b}) \propto 
\left[\left(g_L^b\right)^2 + \left(g_R^b\right)^2 \right]\,,\ \ 
A^b_{FB} \propto  
\frac{ \left(g_R^b\right)^2 - \left(g_L^b\right)^2}
{ \left(g_R^b\right)^2 + \left(g_L^b\right)^2} 
\,,
\label{eq:width-asymm}
\eeq
with $g_R^b = 0.077$ and $g_L^b = -0.423$ within the SM. The
partial width $\Gamma(Z\to b\bar{b})$ can be pushed upward by changes
in either or both of $g_{L,R}^b$; however, the upward pull on
$A^b_{FB}$ preferentially chooses a change in $g_R^b$. This change
must be such that $|g_R^b + \delta g_R^b|^2$ is marginally higher than
$(g_R^b)^2$, and so $\delta g_R^b$ must either be positive and small,
or negative and large.  It may seem that analogous solutions with
large and negative $\delta g_L^b$ (so that the sign of $g_L^b$ is
reversed without changing its magnitude appreciably) should also be
admissible.  Indeed, this is true as far as the $Z$-peak observables
are concerned.  However, away from the $Z$-peak, such a switch would
essentially reverse the sign\footnote{Away from the $Z$-peak, the
  domimant contribution to $A_{FB}$ accrues from the interference
  between the photon and $Z$-mediated amplitudes.} of $A_{FB}(e^+ e^-
\to b \bar b)$ and, hence, run afoul of the
data~\cite{Choudhury:2001hs}. It is intriguing to note that such
considerations do not choose between the two solutions of
Eq.~(\ref{gl_gr_fit})~\cite{Choudhury:2001hs}. It is obvious, though,
that if the shifts $\delta g_{L,R}^b$ come only from perturbative
quantum corrections, then the first solution would be much easier to achieve
than the second.

The strongest phenomenological constraints on NP scenarios arise,
typically, from flavour physics, especially from processes involving the
first two families. This had prompted, over the years, many
constructions wherein the coupling of the NP sector to the SM fermions
is not flavour democratic, but is preferential to the third 
generation. Of particular interest in this context are scenarios that
proclaim the Higgs to be a condensate effecting a dynamic breaking of
electroweak symmetry rather than a fundamental
scalar \cite{tt_condensate}, or models with extensions 
of the gauge group associated with electroweak symmetry \cite{ununif}.
Other examples of models that envisage a special role for heavy 
fermions include models with extra space-time dimensions
\cite{acd_ued,ued_others,Barbieri_ed}, and models where the electroweak 
symmetry is broken in a nonlinear way \cite{Larios:1996fa}, including the 
Little Higgs models \cite{lh0}. A still different class of possibilities is
afforded by the hypotheses where the SM is augmented by colour-triplet or
colour-sextet scalars that have Yukawa couplings with the third generation
\cite{diquark}. 

With each such NP scenario being unique in certain respects, it is
useful to concentrate on the essential aspects, rather than dwell on the 
specifics. In particular, if the NP sector is heavy, integrating
it out would leave us with new operators in the effective low-energy
theory. Moreover, if the NP sector couples preferentially with the
third generation, these would primarily be four-fermion
operators (and, perhaps, anomalous magnetic moment like operators)
involving third generation currents with undetermined Wilson
coefficients that have to be matched with the full NP.
Ref.~\cite{cgk2012}, for example, considered the possibility of such
operators explaining certain tensions in the B-physics sector.
In this paper, we adopt a similar stance and investigate
the implications of such an effective theory for the $Z$-peak
observables, including $R_b$, $A^b_{FB}$ and $A^\tau_{FB}$
and whether some of these
operators could possibly ameliorate the aforementioned discrepancies.
While it might seem that, given the large number of operators
available, it would always be possible 
to find a set that ``solves'' the problem, it turns out that, in reality, 
only a subset can play the
requisite role. Furthermore, a large Wilson coefficient for any such
operator would lead to tell-tale signatures at the LHC, thereby
offering us falsifiability of the ansatz.

This paper is organized as follows. In Sec.~\ref{sec:new-operators},
we introduce new effective dimension-6 operators involving only the
third generation fermions. As our aim is to enhance $\delta g^b_R$, we
might expect that operators involving right-chiral fields would be
more suitable for our purpose, and that indeed turns out to be the
case. We also delineate  the region allowed by the $Z$-peak observables 
in the parameter space of the new operators.  In
Sec.~\ref{sec:LHC_constraints}, we discuss some of the possible
signals at the LHC that should show an unambiguous signature of such
new physics. We summarize and conclude in the last Section. Some
calculational details are relegated to the Appendix.

\section{New operators}
\label{sec:new-operators}

As we are interested essentially in $b$-sector observables, we begin by
introducing generic four-fermion operators involving the $b$ quark,
given by
\be
\frac{\xi}{\Lambda^2} \left[ \bar{f} \gamma_\mu (v_f + a_f\gamma_5) f\right]\,
\left[\bar{b} \gamma^\mu (v_b + a_b\gamma_5) b\right]\,,
\label{eq:O_fb_gen}
\ee
where $\xi$ is a dimensionless number which is \textit{a priori}
undetermined and can only be fixed with a knowledge of the full
theory. $\Lambda$ is the scale up to which the effective theory is
valid, and is essentially the scale of NP.  The identity of $f$ is
undetermined at this point. It is obvious, though, that low energy
constraints on such an operator are the least severe if $f$ is a third
generation fermion. For example, if $m_f < m_b$, we would need 
$\xi / \Lambda^2 \ll \alpha / M_\Upsilon^2$
so as not to run afoul of $\Upsilon(nS)$ decays.
The SM decay is an electromagnetic one, and the width is given by
\cite{sanchis-lozano}
\[
\Gamma_{\Upsilon(1S)\to \ell\ell} = 4\, \alpha^2 \, Q_b^2 \, 
M_{\Upsilon}^{-2} \, |R(0)|^2 \, 
(1+2x) \, \sqrt{1-4x} \,,
\]
where $x = M_\ell^2/M_\Upsilon^2$ and $R(0)$ is the radial part of the
non-relativistic wave function at the origin. It might be argued that
such a decay has non-trivial dependences on quantities (such as
$R(0)$) that can only be calculated in a non-perturbative framework
and, thus, the results are model-dependent. It is easy to see, though,
that apart from the comparable numerical factors, the new physics rate
is suppressed by $\xi/\Lambda^2$ compared to $\alpha/M_\Upsilon^2$,
and so the bound quoted here is a very conservative one.  The terms in
Eq.~(\ref{eq:O_fb_gen}) do not exhaust the list of relevant Lorentz
invariant neutral current four-fermion operators.  Scalar
(pseudoscalar) and tensor (pseudotensor) structures are also
admissible possibilities; however, as would be obvious immediately,
the contributions of such operators to the effective $Z b \bar b$
vertex are chirality suppressed\footnote{It might be argued that
    the contribution of such a (pseudo-)scalar term to
    $\Gamma(\Upsilon(1S) \to \ell^+ \ell^-)$ would be chirality true,
    thereby allowing the corresponding Wilson coefficient to be
    large. On the other hand, this would be severely constrained by
    the non-observation of the $\chi_{b0} \to \ell^+ \ell^-$ decay.}

\begin{figure}[!htbp]
{
\vspace*{-10pt}
\hspace*{-40pt}
\epsfxsize=6cm\epsfbox{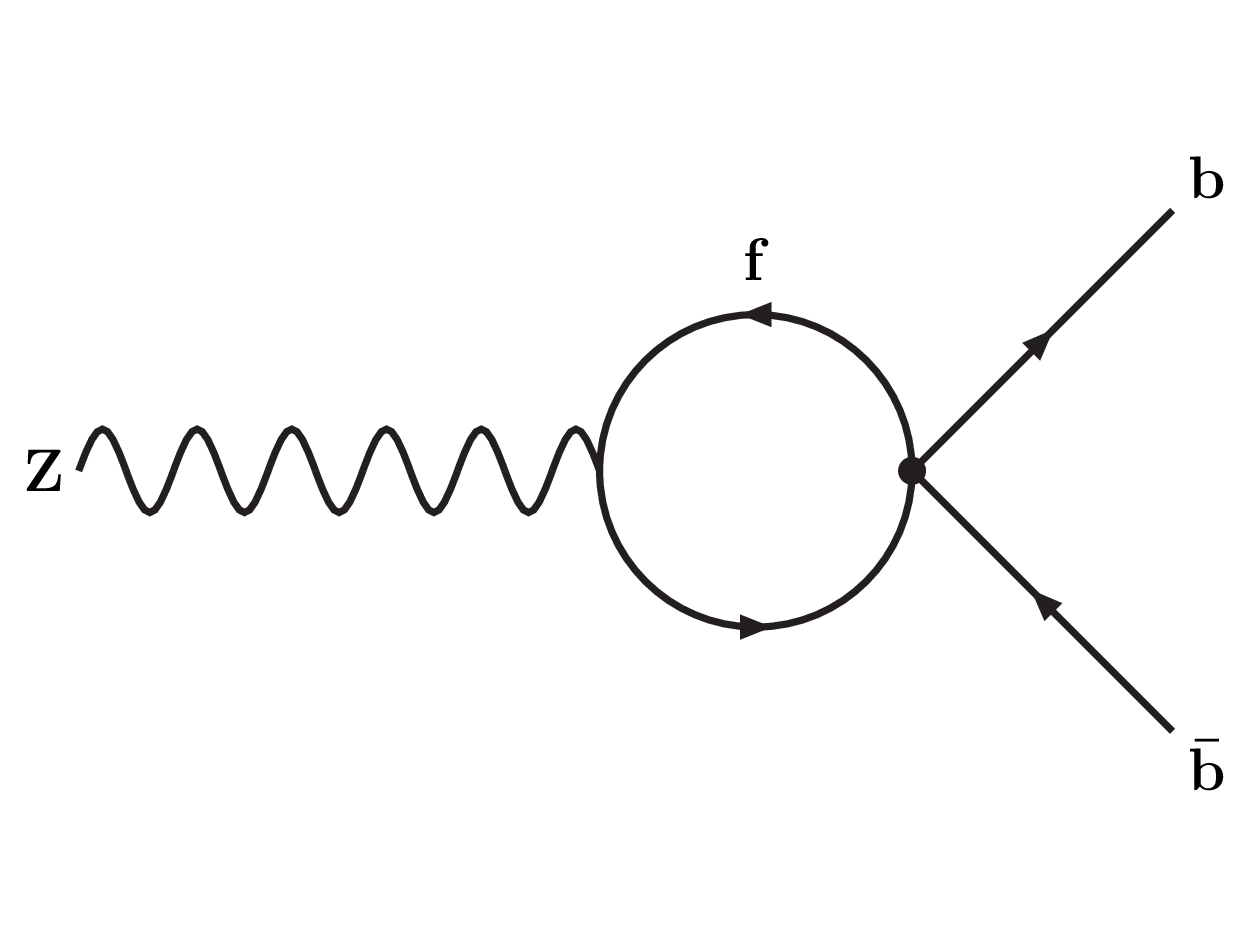}
\vspace*{-10pt}
}
\caption{The effective $Zb\bar{b}$ vertex arising from a single insertion of 
the operator in Eq.~(\ref{eq:O_fb_gen}).}
\label{fig:Zbb_eff}
\end{figure}

The operator of Eq.~(\ref{eq:O_fb_gen}) gives rise to one-loop correction 
to the $Z\to b\bar{b}$ vertex (see Fig.~\ref{fig:Zbb_eff}). Formally, this 
amplitude is quadratically divergent and can be evaluated using a gauge 
invariant prescription such as dimensional regularization. While the infinite 
correction is cancelled by introducing appropriate 
counterterms\footnote{Although this might seem strange given the 
 higher-dimensional nature of the interaction term, note that the calculation 
 fully conforms to the spirit of effective field theories.}, 
 the finite part of the correction to the $Z b \bar b$ vertex is given by
\be
\delta g_L^b 
\; = \;
\left(\dfrac{v_b - a_b}{2}\right) \;
\dfrac{N_C\,\xi}{4\,\pi^2\,\Lambda^2} \;
{\cal J}
\quad \quad ; \quad \quad
\delta g_R^b 
\; = \; 
\left(\dfrac{v_b + a_b}{2}\right) \;
\dfrac{N_C\,\xi}{4\,\pi^2\,\Lambda^2} \;
{\cal J},
\label{eq:delta_gLbgRb}
\ee
where $N_C=3(1)$ if $f$ is a quark (lepton) and ${\cal J} \equiv 
{\cal J}(v_f,a_f, m_f,M_Z)$, the expression for which can be found in
the Appendix. It should be appreciated that, had we attempted instead to
calculate the effective $b \bar b \gamma$ vertex, the very form
of the corresponding ${\cal J}$ would have ensured that the charge radius
does not receive any correction. This, of course, is a consequence of
gauge invariance and has been ensured by our use of dimensional
regularization rather than a naive momentum cutoff \footnote{Note that a
naive application of a cutoff regularization would have given rise
to leading corrections being independent of $\Lambda$ rather than 
being suppressed as $(m^2 / \Lambda^2) \, \ln (m^2 / \Lambda^2)$, with the 
consequence that a smaller $\xi$ would be required. Although such a dependence
of the corrections would have been expected in a scalar theory, it is clearly 
not gauge invariant and, hence, inapplicable in the current context.}. 
If the scale $\Lambda$ of new physics is to be substantially
larger than the electroweak 
symmmetry breaking scale (as the absence of any new resonances 
at the LHC seems to suggest), the four-fermion operators need to respect the full 
$SU(2)_L \otimes U(1)_Y$ symmetry. This is a further restriction on the 
generic operators of Eq.~(\ref{eq:O_fb_gen}). As we need 
$\delta g_R^b \gg \delta g_L^b$, it stands to reason that the said operator 
should involve the $b_R$ field rather than $b_L$. One of the simplest such 
operators is given by
\be
{\cal O}_{RR}^t 
\; = \; 
\frac{\xi}{\Lambda^2} (\bar{t}_R\gamma_\mu t_R)(\bar{b}_R\gamma^\mu b_R)\,.
\label{eq:O_tb_RR}
\ee
{\em i.e.} with the choice $v_b = a_b = v_t = a_t = \frac{1}{2}$.

In the above, we have deliberately neglected the possibility of quark
mixing. Since these operators were presumably generated well above the
electroweak scale, it is likely that they were generated in the weak
basis instead. If the starting point be indeed so, after the symmetry
breaking, the operators need to be re-expressed in terms of mass
eigenstates through a CKM-type rotation~\cite{cgk2012}. This would,
then, generate a plethora of new operators.
The corresponding Wilson coefficients would be constrained by several
B-physics observables such as the mass differences $\Delta M_d$ and
$\Delta M_s$, and the CP violating phases $\beta$ and $\beta_s$. We
apply the principle of Occam's razor and refrain from considering the
entire range of such new operators, restricting ourselves to considering
the operator ${\cal O}_{RR}^t$ only. Note that, apart from 
the phenomenological advantages, an operator such as ${\cal O}_{RR}^t$
is typically less suppressed than others\footnote{Here we discount possible
four-top operators as they are not germane to the issue at hand.}
in scenarios wherein the electroweak symmetry is broken in 
a nonlinear fashion~\cite{Larios:1996fa}.

\begin{figure}[!htbp]
{
\epsfxsize=8cm\epsfbox{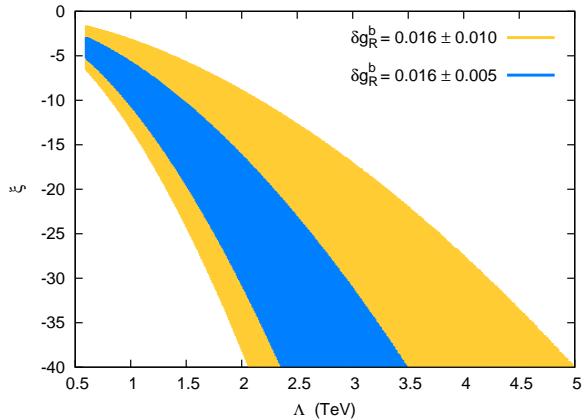}
}
\caption{The allowed region in the $\xi$-$\Lambda$ plane that is
consistent with the observed values of $R_b$ and $A^b_{FB}$.}
\label{fig:xi_vs_Lambda}
\end{figure}

Eq.~(\ref{eq:delta_gLbgRb}) immediately gives $\delta g_L$ = 0 and
$\delta g_R \neq$ 0. The region in the $\xi-\Lambda$ plane that
generates the required $\delta g_R$ (as in Eq.~(\ref{gl_gr_fit})) is shown 
in Fig.~\ref{fig:xi_vs_Lambda}. Requiring that the coupling $\xi$ be
perturbative, at least at the TeV scale, means that only the $\delta
g_R > 0$ solution proposed by Ref.~\cite{batell} is realised \footnote{
One might set the perturbative limit at $\xi\sim {\cal O}(10)$, coming from 
the condition $\xi^2/16\pi^2 < 1$ for higher-order processes in the full 
theory. This is satisfied for $\Lambda\sim {\cal O}(1~{\rm TeV})$, only for the 
$\delta g_R > 0$ solution and not the other one.}.
There is a caveat, though.  The analysis of Ref.~\cite{batell}   was 
performed treating both $\delta g^b_R$ and $\delta g^b_L$ as free parameters, 
whereas invoking ${\cal O}^t_{RR}$ necessarily implies that $\delta g^b_L= 0$.  
In a strict sense, the fit would be different in the two cases.  
However, quantitatively, the $1\sigma$ (or $2\sigma$) allowed regions in the 
two cases are not too different. Indeed, the required $\delta g^b_L$ can be 
generated by positing, in addition, a ${\cal O}_{LL}^t$ with a Wilson 
coefficient much smaller than $\xi$. This, though, would be tantamount to 
invoking two new operators to explain two discrepancies, and, hence, we 
desist from exploring this alternative any further.

It is obvious that the operator in Eq.~(\ref{eq:O_tb_RR}) also modifies the 
$Z t \bar t$ coupling, with the $b$ now in the loop. However probing this 
effect presents a bigger challenge. Even at an $e^+e^-$ collider, $t \bar t$ 
production is dominated by the photon mediated amplitude with 
$e^+e^- \to Z^* \to t \bar t$ making a small contribution. Hence one needs to 
consider more complex processes. We shall return to this discussion in the 
next section.

\subsection{Other operator choices}

As discussed in Sec.~\ref{sec:introduction}, apart from the $b$-sector, some 
minor discrepancies also exist in the $\tau$-sector in the LEP data. 
One may, therefore, contemplate the introduction of an operator
$O_{RR}^{\tau}$ involving $\tau$s and $b$s analogous to $O_{RR}^t$
above, in the hope that the two sets of dispcrepancies could perhaps be
simultaneously explained. However, note that for $O_{RR}^{\tau}$, $N_C$
= 1 for $\delta g_L^{b}$ and $\delta g_R^{b}$, but for the corresponding
corrections to $g_L^{\tau}$ and $g_R^{\tau}$, $N_C$ = 3. Thus, in
general, the corrections to the $Z \tau^+ \tau^-$ couplings will be
larger than those to $Z b \bar b$ couplings\footnote{Although the
correction term also carries a   dependence on the mass of the fermion
in the loop, the difference between $m_{\tau}$ and $m_b$ is small and
cannot entirely offset the difference due to the colour factor.}. On
the other hand, the disagreements between data and SM predictions are
smaller in the case of the $\tau$ observables. Hence, with
$O_{RR}^{\tau}$ alone, it is not possible to simultaeously generate the
requisite corrections to all of $g_L^{b}$, $g_R^{b}$,
$g_L^{\tau}$ and $g_R^{\tau}$. If one were to additionally consider
$O_{RL}^{\tau}$, $O_{LR}^{\tau}$ and $O_{LL}^{\tau}$ as well, it 
is indeed possible to arrange a conspiracy between the coefficients of the 
various operators such that the observed values of the all couplings are
obtained simultaneously. An easier path to such an explanation is
offered by invoking a (set of) $\bar \tau \tau \bar t t $ operators
alongwith $O_{RR}^{t}$. This has the advantage of not upsetting any
other low-energy observable to a significant degree. On the other hand,
it is a construction that is barely testable in current experiments.

A much more intriguing possibility is offered where $f$ 
(in Eq.~(\ref{eq:O_fb_gen})) is an exotic fermion. Clearly, few constraints 
apply to such operators, and it is much easier to arrange for the requisite 
shifts in $g^b_{L,R}$ as long as $f$ itself does couple to the $Z$. 
This is eminently possible, as for example in supersymmetric or 
extra-dimensional extensions of the SM. While many different choices for $f$
are possible (as long as it is heavy enough not to have been found at the 
Tevatron or the LHC), a particularly interesting choice is that of $f$ being 
the dark matter (DM) candidate itself. The tantalizing indications, 
over the years, for the existence of a DM particle (whether it be from 
cosmological data fitting, indirect evidence from satellite-based observations 
or direct earth-bound experiments), in the absence of actual discovery, 
has led to much speculation about 
its nature. It has been realized of late that, quite apart from dedicated 
DM search experiments, collider experiments can provide substantial 
information about the DM sector. Indeed, given the complete absence of any 
information, even dedicated DM searches only parametrize its interactions 
with matter through effective operators as in Eq.~(\ref{eq:O_fb_gen}). 
The very same operators would also lead to DM pair production (in association 
with visible objects) at colliders. Thus, an excess in such channels (with 
the DM pair providing missing momentum) over the SM  expectations would 
constitute a signal while a lack thereof would constrain 
the said interactions~\cite{Goodman:2010yf, Fox:2011pm, Goodman:2010ku, Fox:2012ee}.

The situation becomes particularly interesting if the DM particle
couples to the SM sector preferentially through the third generation
fermions~\cite{Jackson:2009kg,Cheung:2010zf,Bhattacherjee:2012ch,Lin:2013sca}.
Direct detection experiments would be rendered rather ineffectual.
Even satellite-based indirect detection experiments would have reduced
sensitivity. Although collider experiments too would suffer, the
suppression in the cross-section is not that extreme. Aided by the
possibility of tagging heavy flavours, LHC experiments would have the
highest sensitivity (amongst all currently operating ones) to such
operators~\cite{Bhattacherjee:2012ch}.  Given this, it is worthwhile
to consider this possibility as well.  The formalism being identical
to that we have delineated above, the results would only depend on the
choices\footnote{It must be remembered though that if the DM is a
  Majorana fermion, it may not have a vector-like coupling to the $Z$,
  whereas an axial coupling is allowed.}  for the DM couplings to the
$b$-current as well as to the $Z$. And finally, while scalar DM is
also a possibility, and may couple to both the $Z$ as well as to a
$b$-current, the corresponding corrections to the effective $Z b \bar
b$ vertex would have a Lorentz structure that does not readily
translate to a discernible shift in $A^b_{FB}$.
  
\section{$O_{RR}^t$ at the LHC}
\label{sec:LHC_constraints}
In the last section, we saw that the low energy constraints on the
operator $O_{RR}^t$ (or analogous ones) are not strong enough to call
into question a possible role for it in the explanation of the anomaly
in the $Z b \bar b$ vertex. Thus, the only theatre for studying such
an operator is provided by colliders. Although $O_{RR}^t$ also
engenders changes in the $Z t \bar t$ vertex analogous to those
wrought for the $Z b \bar b$ one, such a change is of little relevance
either at the LHC, or even at a linear collider\footnote{Even the best
  sensitivity, provided by a high-luminosity $t \bar t$ threshold scan
  at the linear collider is not adequate to probe the required values
  of $\xi$.}. 
And as we have already argued, loops induced by such
operators do not generate any corrections to the electric or colour
charge radii of the fermions.  Although anomalous (chromo-)magnetic
moments are indeed generated, once again, these are of little
immediate concern as the change in, $ gg \to t \bar t$ is hardly
discernible.

\begin{figure}[!htbp]
{
\epsfxsize=8cm\epsfbox{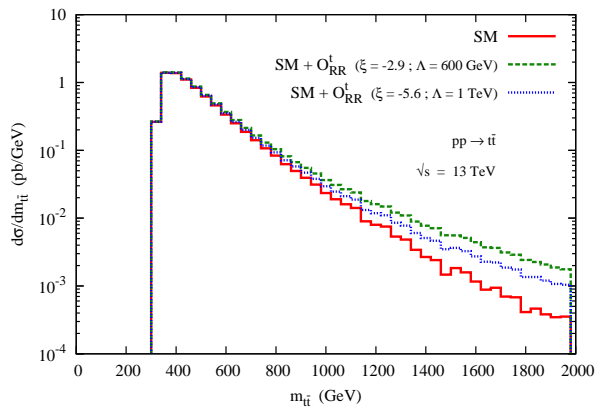}
}
\caption{The $m_{tt}$ distribution of the cross section for 
$pp \to t \bar t$ at $\sqrt{s}$ = 13 TeV, in the presence of an 
anomalous $b \bar b \to t \bar t$ contribution driven by 
$O_{RR}^t$. Included is only the LO cross section computed with the 
CTEQ6L distributions.}
\label{fig:bb_to_tt}
\end{figure}

There, though, is a tree level subprocess that could receive a large
contribution from $O_{RR}^t$, namely $b \bar b \to t \bar t$.  Despite
the smallness of the $b$-flux within the proton, the additional
contribution to the cross-section, at $\sqrt{s} = 13$ TeV, can be as
large as $\sim 10\%$ for values of $\xi / \Lambda^2$ required to
reproduce the correct $\delta g_R^b$ (see
Fig.\ref{fig:bb_to_tt}). While this might seem very promising in view
of the accuracy in the $t \bar t$ cross-section measurement
(especially in the dilepton channel), note that the theoretical errors
due to higher-order corrections and PDF ambiguities are much
larger. The last mentioned is of particular relevance here as the
$b$-flux is relatively poorly known. One might attempt to exploit the
fact that owing to the higher-dimensional nature of the interaction
term, the corresponding amplitude grows with energy. While this is
certainly true at the subprocess level, the growth of the anomalous
cross-section is muted owing to the rapid fall of the $b$-flux with
Bjorken-$x$. Moreover, reconstruction of $m_{tt}$ is less efficient in
the dilepton channel, whereas the use of the hadronic channels
typically lead to larger experimental uncertainties. Given this
situation, we desist from further consideration of this channel.

\begin{figure}[!htbp]
{
\epsfxsize=5.5cm\epsfbox{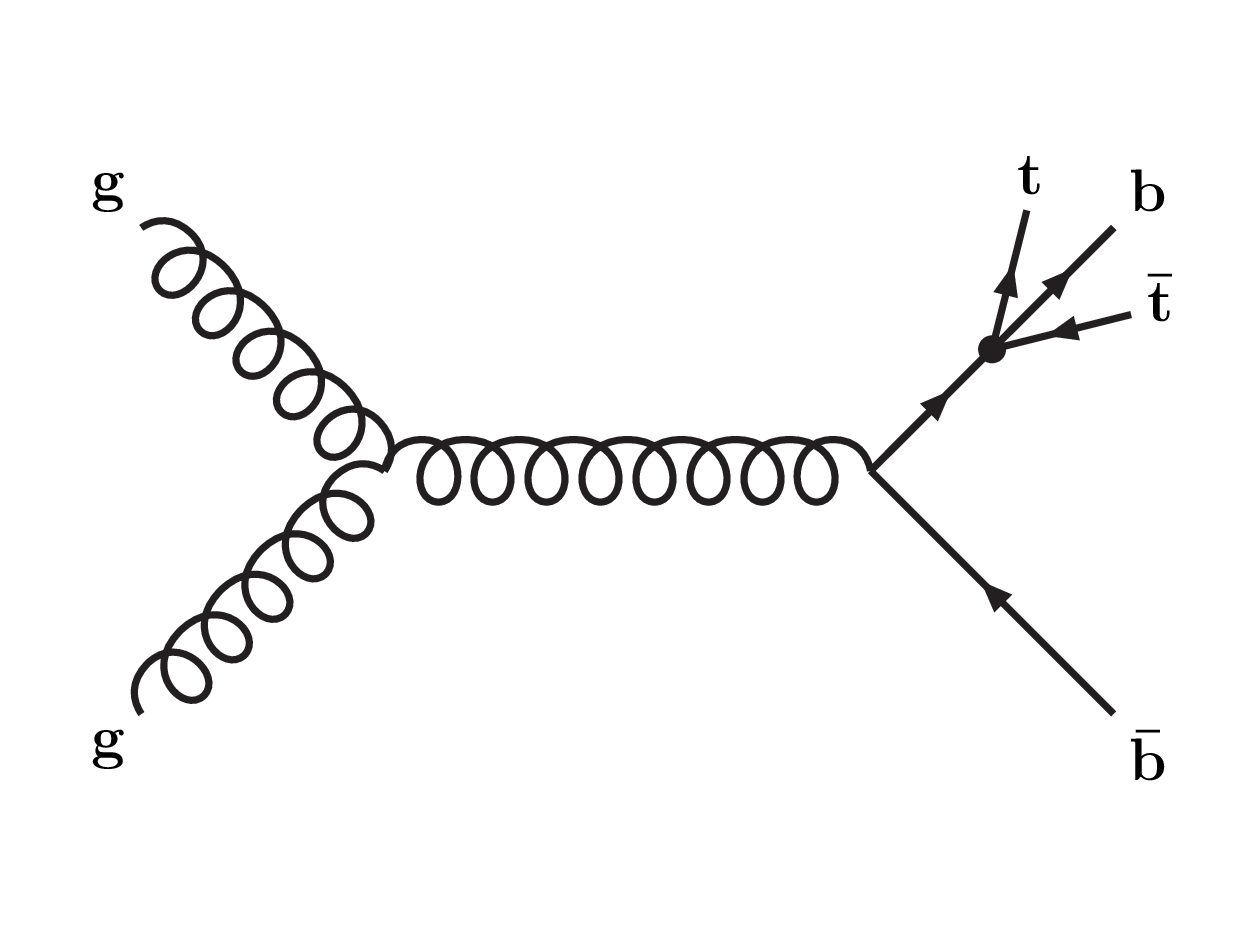}
}
{
\epsfxsize=5.5cm\epsfbox{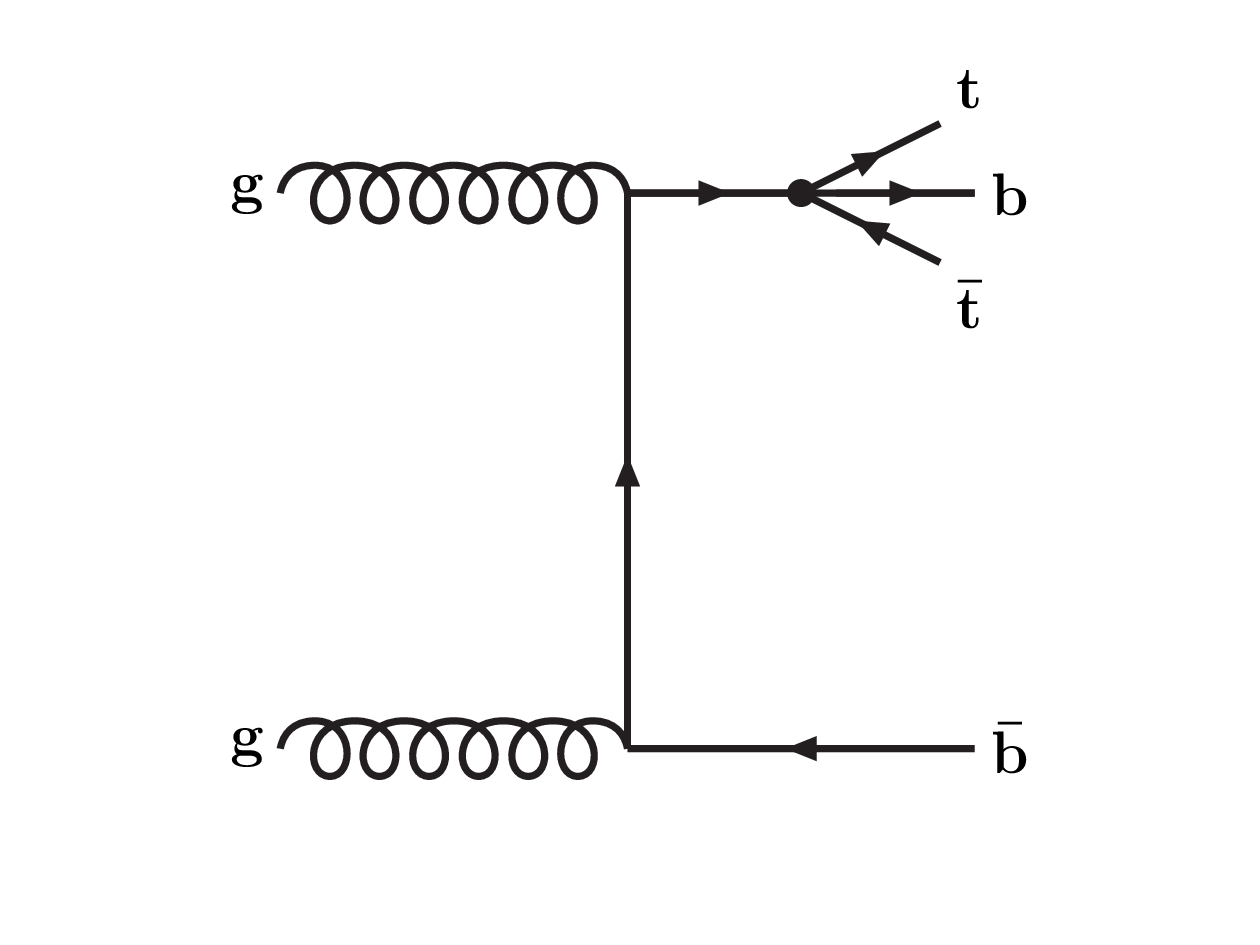}
}
{
\epsfxsize=5.5cm\epsfbox{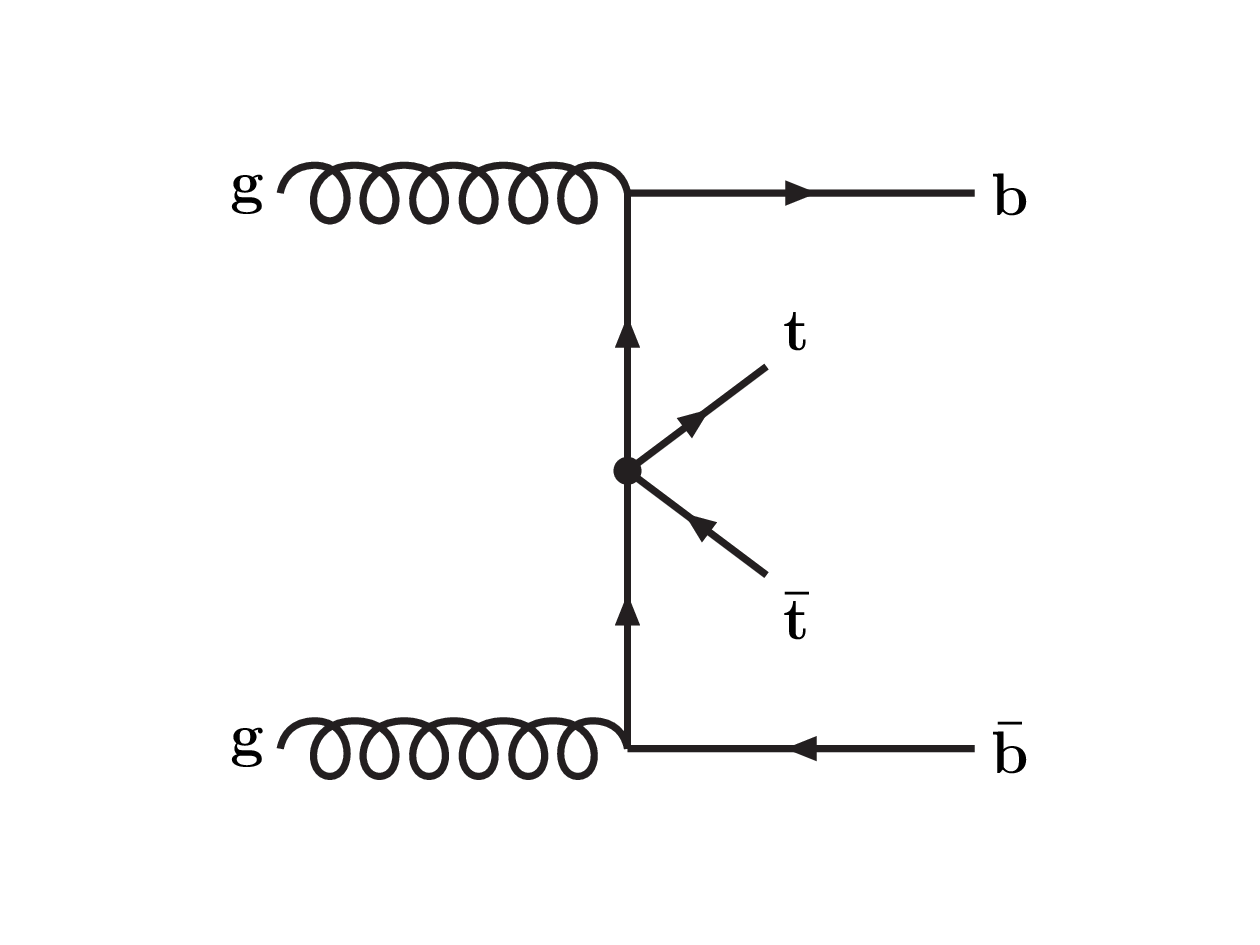}
}
{
\epsfxsize=5.5cm\epsfbox{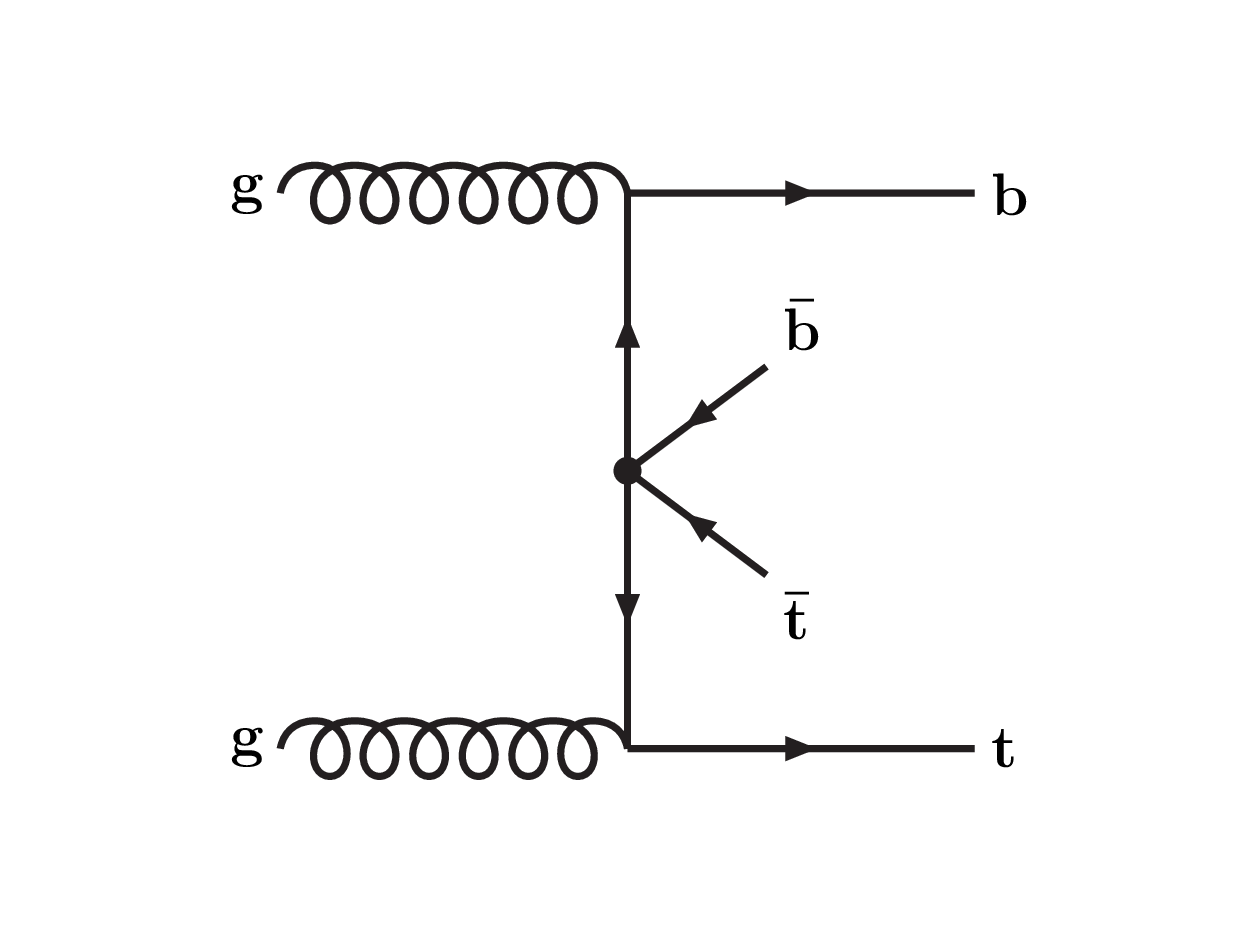}
}
{
\epsfxsize=5.5cm\epsfbox{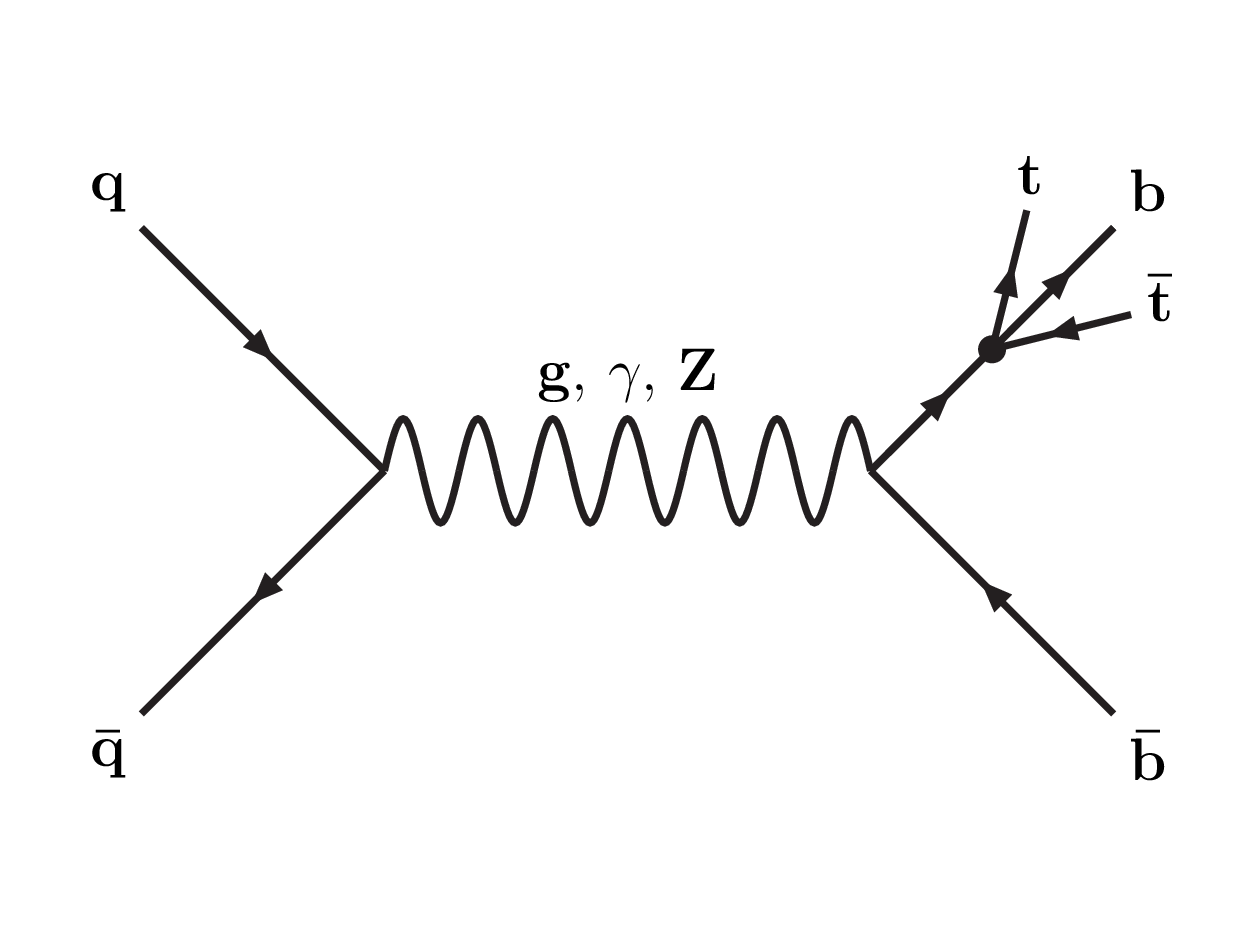}
}
\caption{Some of the new Feynman diagrams that come into play when 
$O_{RR}^t$ is introduced.}
\label{fig:tTbB}
\end{figure}

Instead, we consider the process $p p \to t \bar t b \bar b$.  As
such, this final state is of interest as an SM background for analyses
concerning Higgs production in association with a top-pair where the
Higgs then decays into a bottom pair.  With the introduction of
$O_{RR}^t$, several new diagrams come into play.  Rather than listing
all of them, we illustrate some representative topological classes in
Fig.~\ref{fig:tTbB}.  At the LHC, the gluon-initiated contribution is,
understandably, the dominant one. At first, it might seem that, owing
to a different colour structure, the $O_{RR}^t$ diagrams cannot
interfere with the pure QCD ones. This argument, though, holds only
for those pairs of diagrams wherein the $O_{RR}^t$ vertex is replaced
by a gluon propagator, and not in general. Similarly, the new diagrams
do interfere with the majority of the mixed QCD-electroweak diagrams
in the SM. We incorporate all such potential contributions (including
the subdominant ones) and calculate the cross section through a simple
modification of the CalcHEP~\cite{CalcHEP} software.

For a quantitative assessment, one must impose a minimal set of
acceptance cuts on the final state particles. To this end, we require
that the transverse momentum and the rapidity of the two primary
$b$-jets (i.e., the $b$-jets emanating from 
the primary hard process, rather than the decays of the top) satisfy
\beq
p_T(b) > 50 \, {\rm GeV}\,,   
\qquad 
|\eta(b)| < 2.5\, .
\label{eq:indiv_cuts}
\eeq
To veto $Z$- and Higgs-events (for example, as occasioned from 
$t \bar t Z$ or $t \bar t h$ production), we impose, in addition,
\beq
 M(b, \bar b) \not\in [75, 135]~{\rm GeV}\,.
\label{eq:indiv_cuts_2}
\eeq

\begin{figure}[!htbp]
{
\epsfxsize=8cm\epsfbox{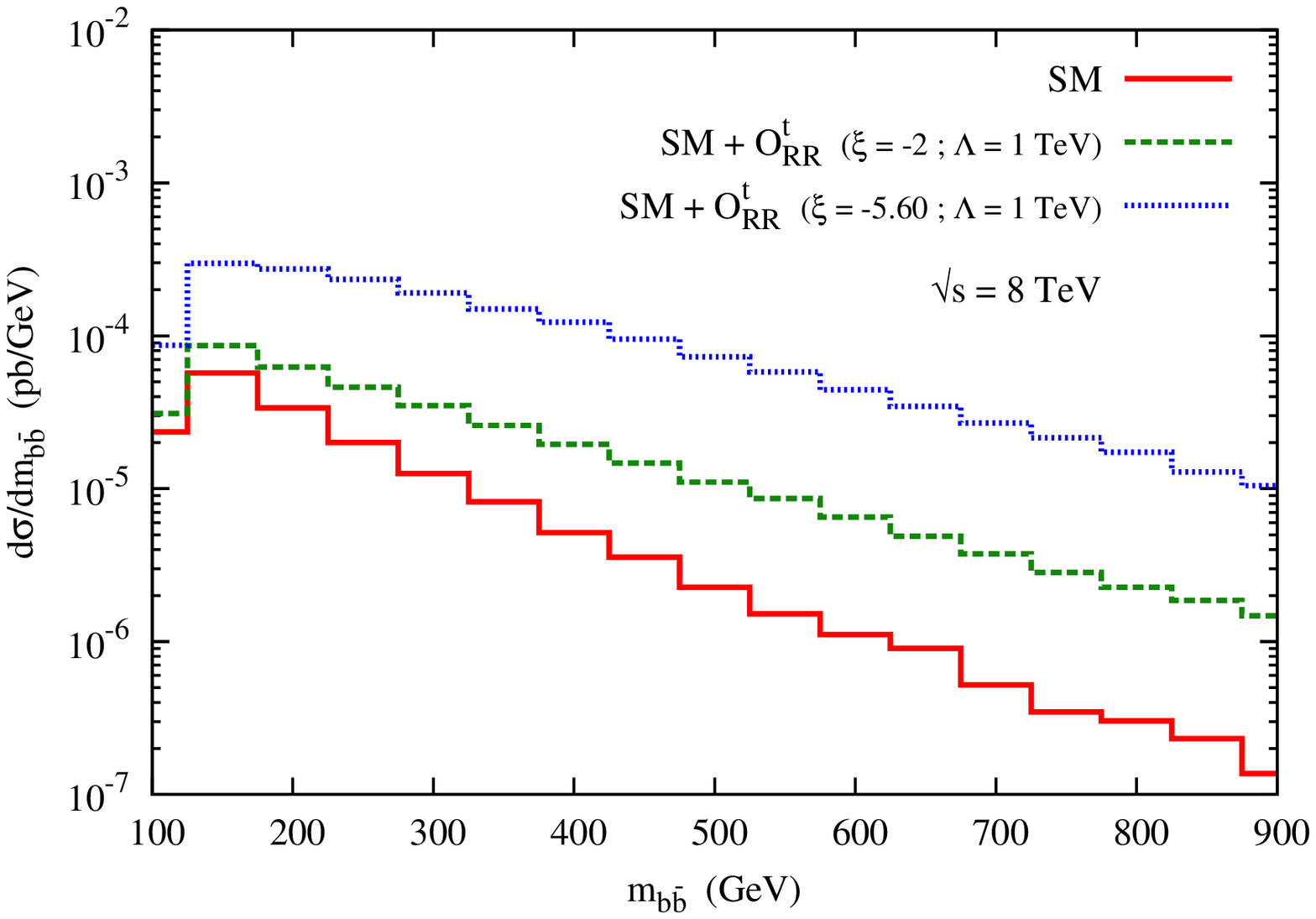}
}
{
\epsfxsize=8cm\epsfbox{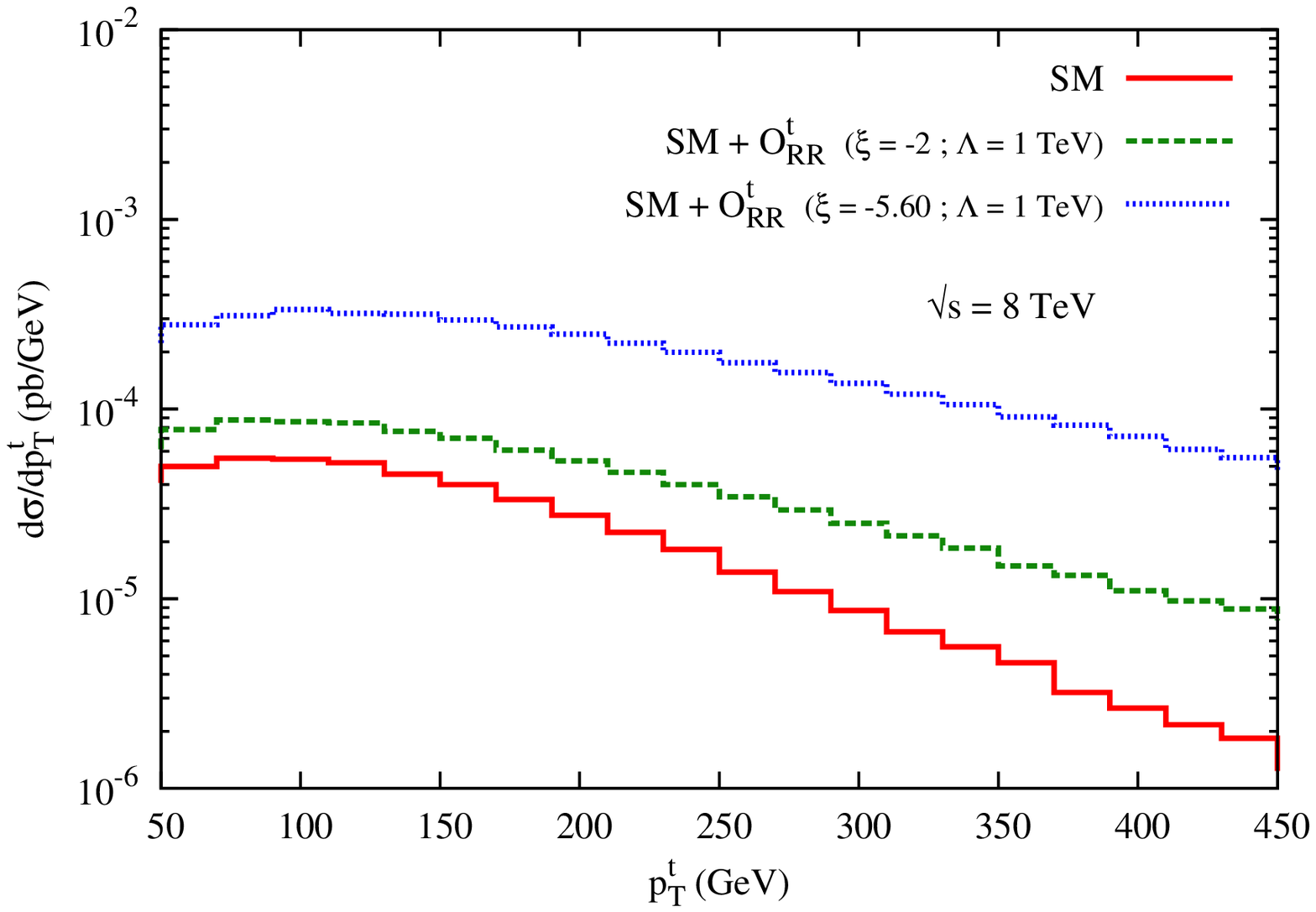}
}
\caption{$pp \to b \bar b t \bar t$ at $\sqrt{s}$ = 8 TeV. 
\textit{Left panel} : 
Invariant mass of the $b \bar b$ system. 
\textit{Right panel} : 
Transverse momentum of the top.}
\label{fig:LHC_8TeV_plots}
\end{figure}

\begin{figure}[!htbp]
{
\epsfxsize=8cm\epsfbox{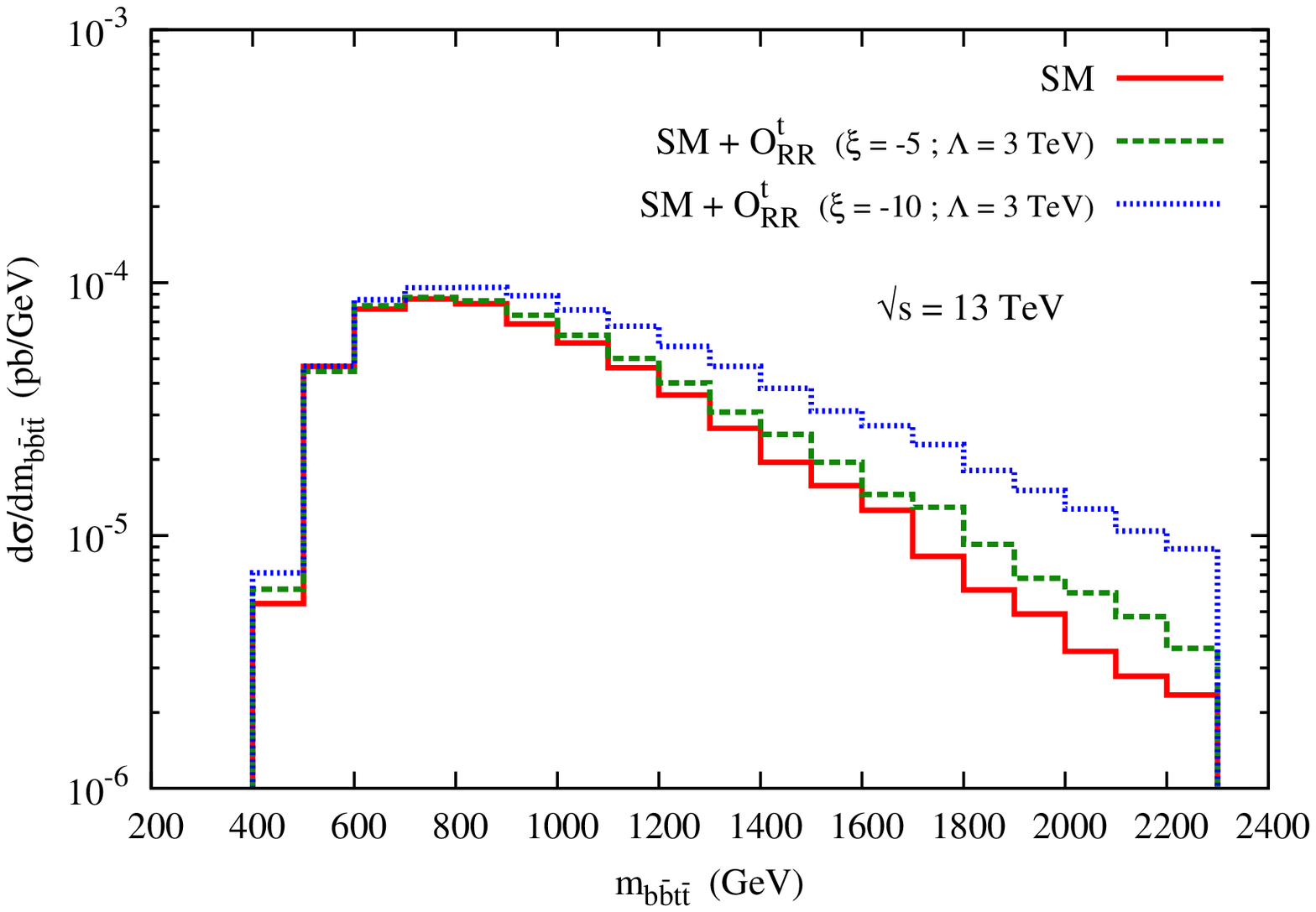}
}
{
\epsfxsize=8cm\epsfbox{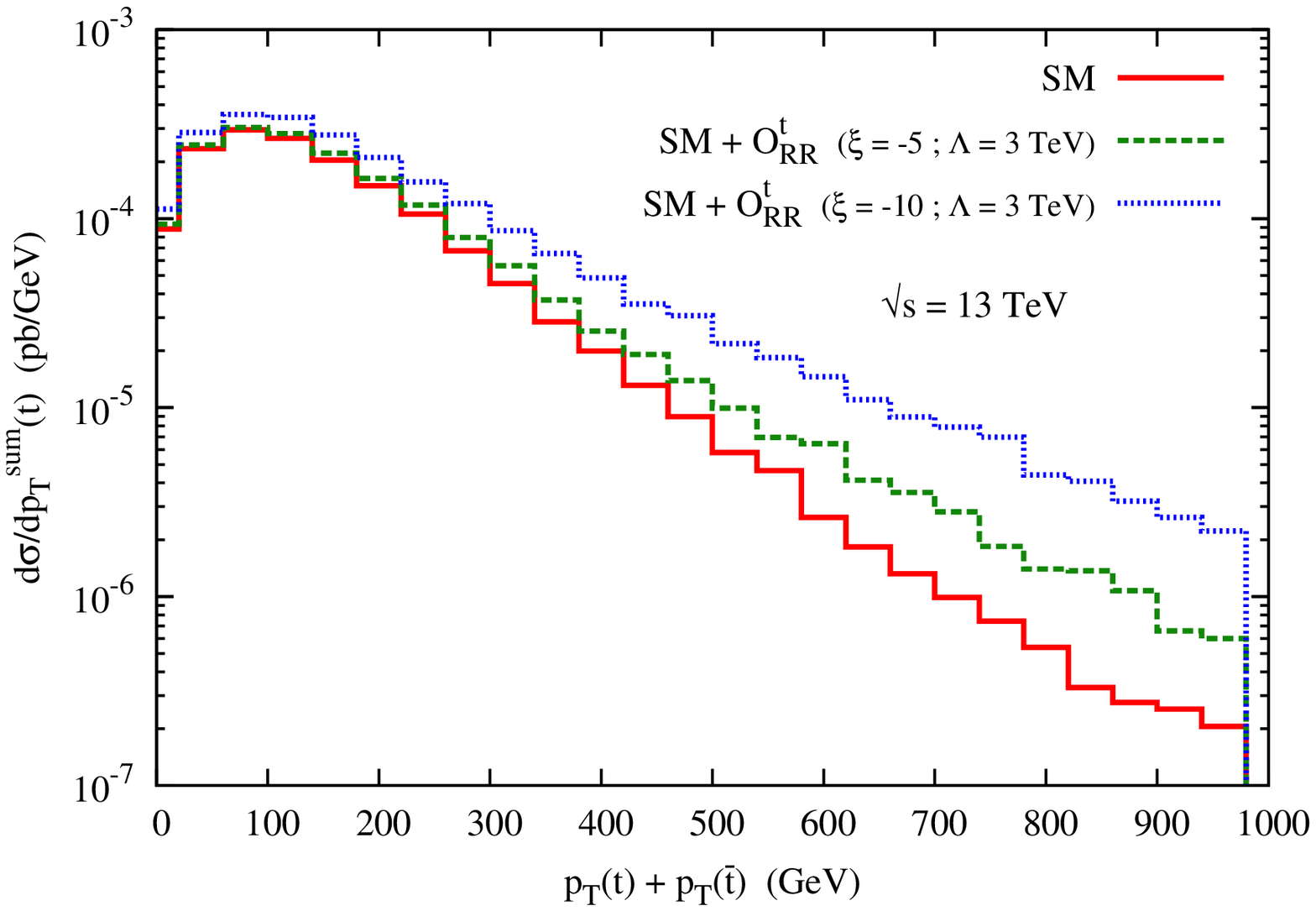}
}
\caption{$pp \to b \bar b t \bar t$ at $\sqrt{s}$ = 13 TeV. 
\textit{Left panel} : 
Invariant mass of the $b \bar b t \bar t$ system. 
\textit{Right panel} : 
Transverse momentum of the $t \bar t$ system.}
\label{fig:LHC_13TeV_plots}
\end{figure}

For a $pp$ collider operating at a centre-of-mass energy of 13 TeV, the
SM prediction for the cross-section for this process as calculated 
using CalcHEP~\cite{CalcHEP} is $\sim$ 60 fb. This could be enhanced by as 
much as an order of magnitude for ($\Lambda,\xi$) values consistent with the 
$Z \to b \bar b$ measurements (see Fig.~\ref{fig:xi_vs_Lambda}). Owing to the 
higher-dimensional nature of the coupling, the excess would, typically, be 
concentrated in phase space regions corresponding to large momentum transfers. 
In Fig.~\ref{fig:LHC_8TeV_plots} and Fig.~\ref{fig:LHC_13TeV_plots}, we show 
some such kinematic distributions. We find that rather 
than require individual particles to be harder or more central, as in 
Eq.(\ref{eq:indiv_cuts}), it is more profitable to impose stronger cuts on 
variables such as those appearing in Fig.~\ref{fig:LHC_8TeV_plots} and 
Fig.~\ref{fig:LHC_13TeV_plots}. 
Apart from the simplistic observables considererd here, in the actual
experimental set-up, the use of more advanced analysis techniques 
will offer additional means for extraction of signal from the
background~\cite{CMS-PAS}.

\vspace*{10pt}

\vspace*{10pt}

Note that the QCD cross-section for the production of a $t \bar t$ pair
alongwith two well separated and hard jets is much larger than the
$\sigma(t \bar t b \bar b)$ that is quoted here. Thus, $b$-tagging is of
prime importance. The corresponding efficiency has a strong dependence
on $p_T(b)$, and thus, requiring it to be very large would lead to a
drastic reduction in signal sizes. On the other hand, the typical values
of $p_T(t/\bar t)$ are not so large as to warrant worries 
pertaining to the identification of highly boosted tops. Thus,
stiffening the cuts on the top momenta would seem to be called for. 
Reconstructing a top, however, is associated with certain limitations.
With the additional bottom pair introducing further combinatoric
ambiguities, the errors would be amplified to an extent. 
Note, though, that owing to its four-fermi nature, 
the signal events would tend to concentrate at higher values of 
$M(b, \bar b)$ where the $b$-jets emanate from the hard process. 
Thus, requiring that for at least one pairing, $M(b, \bar b)$ is 
much larger than the cut of Eq.~(\ref{eq:indiv_cuts_2}) stipulates,
would enhance the signal to noise
ratio~\cite{Degrande:2010kt,Bhattacherjee:2012ch}. 
Indeed, given
that the NP cross-sections are significantly large, the nominal 
luminosity expected for the 13 TeV run of the LHC would be enough for a 
discovery even after accounting for the branching fractions,
$b$-tagging efficiencies, combinatoric ambiguities as well as detector
acceptance and efficiencies for a $\Lambda$ near 3 TeV.
This contention is supported by the 
detailed simulation of Ref.\cite{Bhattacherjee:2012ch}, where
production of Dark Matter particles in association with a top pair has
been considered. Although the final state is different 
($t \bar t + \slashed E_T$), the analysis is similar; the absence of the
missing transverse momentum is amply compensated for by the two hard
$b$-jets. Were one to admit smaller values of $\Lambda \sim$ 1 TeV, 
large deviations from the SM would be expected even in the 8 TeV LHC data 
(see Fig.\ref{fig:LHC_8TeV_plots}).  This mode, thus, is potentially the 
best bet for a direct confirmation of such an ansatz as presented here.

We refrain, though, from using this study to extract information on 
$\xi/\Lambda$. For one, we have not taken into account 
the complexities of event reconstruction for this
final state in the LHC environment. Furthermore, the theoretical predictions
are only the leading order ones. Nonetheless, it is suggestive of a method
that could be used to further investigate a LEP/SLD anomaly at the LHC, where
a direct repetition of the measurement is not possible.

\section{Conclusion}
\label{sec:conclusion}
We have tried to gain some insight into the possible structure of NP at
the TeV scale that might successfully address the mismatch between
measurements and theoretical predictions of $R_b$ and $A^b_{FB}$. 
We have used a bottom-up approach, not being confined to any specific
model, with the sole assumption being that the NP couples only to the
third generation fermions. While there can be several such operators
with different fermion fields and Lorentz structures, electroweak
precision data and B physics observables already put severe constraints
on the Wilson coefficients of most of these operators. 
The quest for an operator that can resolve the anomalies while being
relatively unconstrained has motivated us to work with one involving
right-chiral top and bottom quark fields. At the same time, other choices 
are also possible, {\em e.g.} one with $b$ quarks and dark matter 
particles that couple to the $Z$. 

The four-fermion operators arise from a more fundamental theory at
the higher scale. We perform our analysis in the spirit of an effective
theory, with a high cut-off at the TeV scale (possibly indicative of the
NP masses). The shifts in the $Zb\bar{b}$ couplings are caused by the
parameters of the full theory, and we can only make the leading-order
estimate of these in the effective theory. It turns out that there is a
significant region in the parameter space that is consistent with the
$R_b$ and $A^b_{FB}$ data, without being in contradiction with other
observables.

Finally, we look for the possible signals of this operator at the LHC.
Altough $b \bar b \to t \bar t$ is the lowest order process that 
features the new coupling, given the experimental as well as theoretical 
uncertainities, the sensitivity is likely to be low.
On the other hand, $pp\to t\bar{t}b\bar{b}$ is far amenable to this task.
We find that several observables would show a clear deviation from the SM, 
thus opening up clear channels to investigate such interactions. 
The results will be eagerly anticipated.

{\em Note added:}
Recently Freitas and Huang \cite{Freitas-Huang_revised} have revised 
their two-loop calculation of $R_b$ and $\sin^2 \theta^{b \bar b}_{\rm eff}$. 
As a result the discrepancy between the SM expectations and the experimental
value has again come down to $1.2\sigma$. If it stands, this result would serve
to restrict the parameter space for the higher-dimensional effective
operators. However, even the remaining part of the parameter space
would still be of great interest in the context of the LHC as well as
the paradigm of non-linear realization of the electroweak symmetry.
Moreover, if the anomaly in $R_b$ indeed disappears after the
inclusion of two-loop effects, then it leaves the heavy quark sector in a
rather intriguing position. It appears now that, for third-generation
quarks, production cross-sections agree with SM predictions but $A_{FB}$
measurements do not. This throws up interesting possibilities and is 
sure to spur futher activity in this area in the near future.
\appendix

\section{Analytic Expressions}
\label{sec:expressions}

\begin{figure}[!htbp]
\hspace*{-40pt}
\epsfxsize=6cm\epsfbox{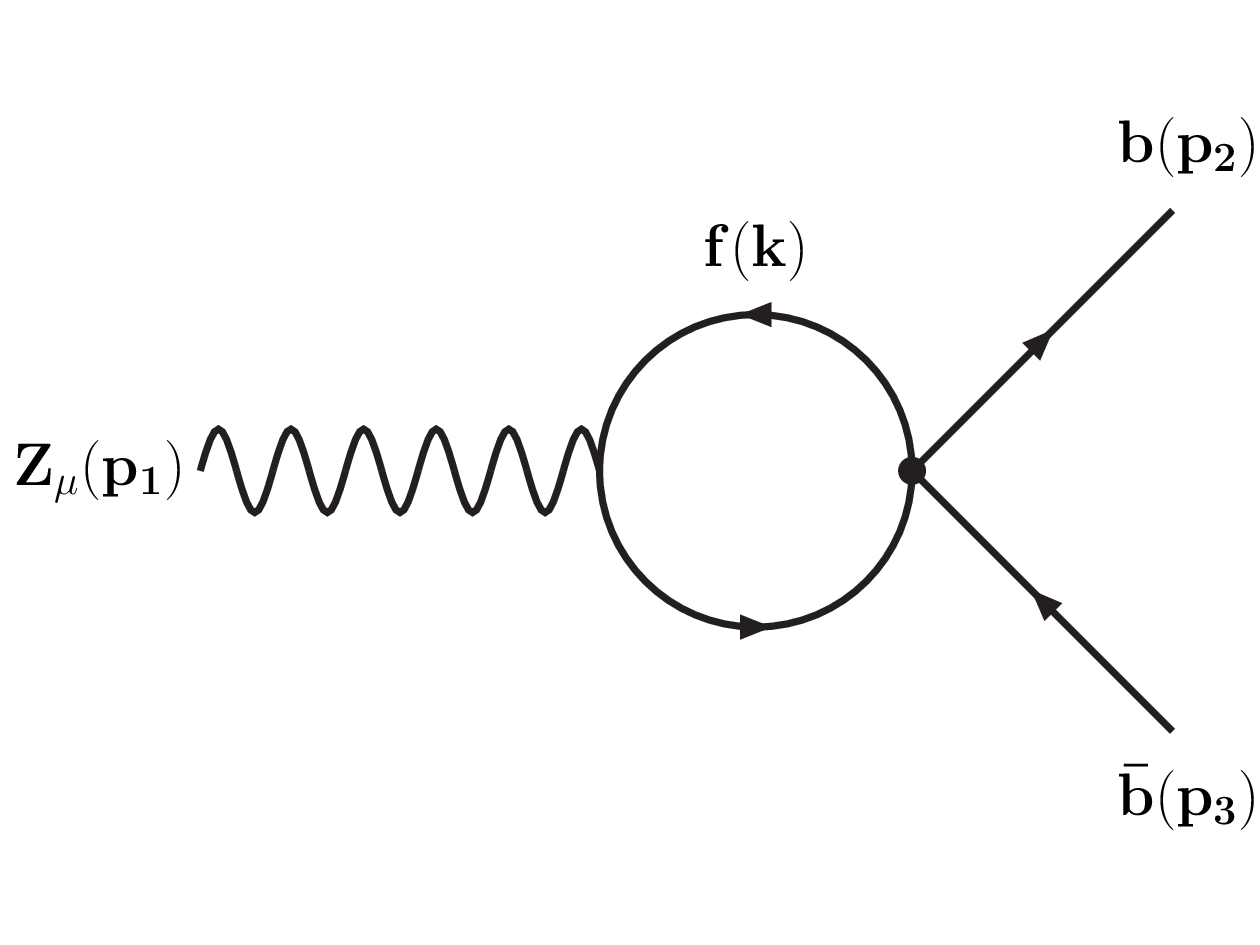}
\vspace*{-10pt}
\caption{One loop correction to the $Z b \bar b$ vertex owing to NP interactions.}
\label{fig:Zbb_eff_labels}
\end{figure}

\noindent
We parametrize the $Z b \bar b$ vertex within the Standard Model 
by 
\begin{equation}
\dfrac{i g}{2 \, \cos\theta_W} \; 
\bar b \, \gamma^{\mu} (\vzb + \azb \gamma_5) \, b \, .
\end{equation}
The one-loop correction to this vertex on account of the 
interaction of eqn.(\ref{eq:O_fb_gen}) is given by the diagram of 
Fig.\ref{fig:Zbb_eff_labels}. 
The expression for the corresponding correction is given by
\beq
\dfrac{g N_C \xi}{2 \cos\theta_W \Lambda^2}
[\bar b \, \gamma_{\alpha} (\vb + \ab \gamma_5) \, b ] 
\cdot \; \Gamma^{\mu\alpha}
\eeq
where
\beq
\Gamma^{\mu\alpha} = 
-
\int \dfrac{d^4k }{(2\pi)^4}
\dfrac{
Tr
\left[
\gamma^{\mu} \, (\vzf + \azf \gamma_5) \, (\kay + \mf) \,
\gamma^{\alpha} \,
(\vf + \af \gamma_5) \, (\kay + \pone + \mf) 
\right]
}
{
(k^2 - \mf^2) \;
\left[(k + p_1)^2 - \mf^2\right]
} \,.
\eeq
The evaluation of this integral is best done by ignoring  
the higher-dimensional nature of the coupling and the possible 
role of $\Lambda$ as a cutoff. Treating ($\xi / \Lambda^2$) 
just as a dimensionful parameter in the theory, we employ 
dimensional regularization and the
 finite part of the correction is given by 
(note that $\chi < 0$ denotes the presence of a threshold)
\begin{equation}
\Gamma^{\mu\alpha}
 =  
\frac{i\,g^{\mu\alpha}}{4\pi^2} \; {\cal J}
\end{equation}
with 
\begin{eqnarray}
 {\cal J} &=& 
\frac{2}{3} \; \aplus \, p_1^2 \; 
\bigg[
\frac{1}{2} \; \ln\left(\frac{\mf^2}{\Lambda^2}\right) \;+\;
\frac{1}{6} \;- 2\; \chi \;-\; \frac{3}{2} \;+\; 
\sqrt{\chi} \, \left(3 + 4 \, \chi \right)
\; \tan^{-1}\left( \frac{1}{2\sqrt{\chi}} \right)
\bigg] \nonumber \\
&& - 
 \mf^2 \; (\aplus-\aminus) \; 
\left[
\ln\left(\frac{\mf^2}{\Lambda^2}\right) \;-\ 2 \; + \; 4\sqrt{\chi} \; 
\tan^{-1}\left( \frac{1}{2\sqrt{\chi}}\right)
\right]\,,
\end{eqnarray}
where
\begin{equation}
 \chi 
 =  \frac{\mf^2}{p_1^2} \, - \, 
\frac14\,,\qquad 
{\cal A}_{\pm}
 =
\vzf \vf \; \pm \; \azf \af\,.
\end{equation}
In other words, on the inclusion of NP, 
\begin{equation}
\vzb \quad \longrightarrow \quad
\vzb \;+\; \vb \dfrac{N_C \, \xi}{4\pi^2\Lambda^2} \; {\cal J}
\qquad \text{;} \qquad 
\azb \quad \longrightarrow \quad
\azb \;+\; \ab \dfrac{N_C \, \xi}{4\pi^2\Lambda^2} \; {\cal J}
\end{equation}
or, in terms of $g_L^b$ and $g_R^b$,
\begin{equation}
g_L^b \quad \longrightarrow \quad 
g_L^b + \left(\dfrac{v_b - a_b}{2}\right) \; 
        \dfrac{N_C\,\xi}{4\,\pi^2\,\Lambda^2} \; {\cal J}
\qquad \text{and} \qquad 
g_R^b \quad \longrightarrow \quad
g_R^b + \left(\dfrac{v_b + a_b}{2}\right) \; 
        \dfrac{N_C\,\xi}{4\,\pi^2\,\Lambda^2} \; {\cal J}
\end{equation} 

{A couple of points need to be noted here. Had we employed a 
naive cut-off regularization instead, we would have encountered a 
quadratic divergence instead of the logarithmic one present in ${\cal J}$. 
This, however, would have been a spurious one occasioned by the facts
that the loop integral is a tensorial one and that the naive cut-off 
regularization does not respect the symmetries of the theory}~\cite{Degrande:2012gr}.
Indeed, the adoption of such a regularization would have induced 
anomalous corrections to the (chromo-)electric charge radius of the 
$b$ and the $t$, thereby violating gauge invariance.
On the other hand, had we used a gauge and Lorentz-invariant prescription such as 
the Pauli-Villars scheme, we would have obtained a term exactly analogous to 
that we already have, achieved though after a much more tedious calculation. 
A further issue relates to our implicit equalization of the renormalization 
scale $\mu_R$ with $\Lambda$. While this choice is a natural one, it is by no means 
the only possible one. Note, though, that the additional term introduced 
by using $\mu_R \neq \Lambda$ is a subdominant one and of little consequence here.


\section*{Acknowledgements}

AK acknowledges support from CSIR, India, and the DRS programme of UGC, India. 
PS would like to acknowledge financial support from NSERC, Canada. 



\end{document}